\documentclass[AMA,STIX1COL]{WileyNJD-v2}
\usepackage{moreverb}
\usepackage{xcolor}
\usepackage{url}

\newcommand\BibTeX{{\rmfamily B\kern-.05em \textsc{i\kern-.025em b}\kern-.08em
T\kern-.1667em\lower.7ex\hbox{E}\kern-.125emX}}

\newcommand{\bfX}{{\bf X}}

\articletype{Article Type}%

\received{<day> <Month>, <year>}
\revised{<day> <Month>, <year>}
\accepted{<day> <Month>, <year>}

\begin{document}

\title{A matching design for augmenting a randomized clinical trial with external control}

\author[1]{Jianghao Li}
\author[1]{Yu Du}
\author[1]{Huayu Liu}
\author[1]{Yanyao Yi*}

\authormark{Li, \textsc{et al}}

\address{\orgdiv{Department of Biometrics}, \orgname{Eli Lilly and Company}, \orgaddress{\state{IN}, \country{United States}}}

\corres{*Yanyao Yi,  Department of Biometrics, Eli Lilly and Company, \email{yi\_yanyao@lilly.com}}

\presentaddress{Lilly Corporate Center, Indianapolis, IN 46225.}

\abstract[Abstract]{The use of information from real world to assess the effectiveness of medical products is becoming increasingly popular and more acceptable by regulatory agencies. According to a strategic real-world evidence framework published by U.S. Food and Drug Administration, a hybrid randomized controlled trial that augments internal control arm with real-world data is a pragmatic approach worth more attention. In this paper, we aim to improve on existing { matching designs for such a hybrid randomized controlled trial}. In particular, we propose to match the entire  concurrent randomized clinical trial (RCT) such that (1) the matched external control subjects used to augment the internal control arm are as comparable as possible to the RCT population, (2) every active treatment arm in an RCT with multiple treatments is compared with the same control { group}, and (3) matching can be conducted and the matched set locked before treatment unblinding to better maintain the data integrity. Besides a weighted estimator, we also introduce a bootstrap method to obtain its variance estimation. The finite sample performance of the proposed method is evaluated by simulations based on data from a real clinical trial.}

\keywords{external control, matching without replacement, propensity score, real-world data, real-world evidence}

\maketitle

\section{Introduction}\label{sec1}
Although randomized clinical trial (RCT) still plays the central role in studying the casual effect of a treatment compared with a placebo, there are increasing explorations on leveraging external information from real world to modernize clinical trial design and accelerate medicinal product development. The 21st Century Cures Act, signed into law in December 2016, placed additional focus on using real-world data (RWD) to generate real-world evidence (RWE) to support regulatory decisions about the effectiveness of medical products in addition to its traditional use for evaluation of safety. { This trend} can also be seen from the extensive effort made by FDA on development of RWE and complex innovative design (CID) programs\cite{FDARWEweb, FDACIDweb}, for which a series of  guidance documents have been published. According to a strategic RWE framework \cite{FDARWEprogram} published by FDA, a potential use of RWD to support effectiveness is to design hybrid randomized controlled trials which combine internal control arm with patients' data from real world.

Recent years have seen the exploding trend of publications on supplementation of the concurrent control of an RCT by external data. However, research on this topic could date back to 1976, where Pocock\cite{Pocock1976} discussed the conditions for acceptability of a historical control group and proposed a bias model to combine the estimators from concurrent and external control data. Following then, two workstreams emerged in this field. One is developed under Bayesian framework, where the external control information is integrated into the RCT through the construction of prior distributions, such as, weighted log-likelihood based methods represented by power prior \cite{Chen&Ibrahim2000, Duan2006, Neuenschwander2009, Banbeta2019, Lin2019, Wang2019}, and hierarchical modeling based methods represented by meta-analytic predictive prior \cite{Neuenschwander2010, Schmidli2014, Hupf2021, Liu2021}. The former methods take heterogeneity of concurrent and external controls into account by discounting the degree of borrowing, while the latter methods assume exchangeability and uses a random effect to quantify and account for the data heterogeneity. See the paper by Rosmalen et al. \cite{Rosmalen2018} for a comprehensive review and comparison of the Bayesian methods. 
The other stream follows the Frequentist approach built on casual inference framework. To our best knowledge, unlike the abundance in the publication of Bayesian methods, the Frequentist approach has limited publications on this topic, but many existing techniques developed for observational studies can be leveraged. For example, weighting (e.g., propensity score \cite{Rosenbaum1983, Robins1994, Hirano2001}, entropy balancing score \cite{Hainmueller2012, Zhao2017} and covariate balancing score \cite{Imai2014}), matching \textit{with} replacement \cite{Abadie2006, Abadie2016} based on covariates or propensity score, and matching \textit{without} replacement \cite{Rosenbaum1985, Gu1993, Rosenbaum2020} based on different tools (e.g., distance metrics and propensity score caliper) and algorithms (e.g., network-flow optimization and greedy-algorithm). All these methods intend to remove confounding bias from the observed covariates. Then, by assuming no unmeasured confounding bias, which is similar to the exchangeability assumption under the Bayesian approach, the external control can be utilized to supplement the estimate of the outcome along with the concurrent control in an RCT and hence to improve the estimation efficiency. An enlightening work done by Stuart and Rubin \cite{Stuart2008} proposed a Frequentist approach to obtain matches from multiple control groups. Although this work was primarily developed for observational studies, it could be a potential application to augment an RCT with external control data as mentioned in the paper. Yuan at el. \cite{Yuan2019} adapted Stuart and Rubin's method for the setting of RCT with two arms where the treatment arm has a larger sample size than the concurrent control arm. This setting is not uncommon in that more treatment exposure is preferred by the sponsor than the control especially given the resource constraint or regulatory requirement on treatment exposure. Yuan at el. \cite{Yuan2019} randomly subsampled matches for the concurrent control from the active treatment group, then the rest of the treated patients were matched by external control. This is a reasonable approach given that the nature of RCT (i.e., randomization) guarantees that the concurrent control and active treatment arm are comparable with respect to both measured and unmeasured covariates. They called their method NC (No Concurrent) matching.

In this paper, we propose a method that improves on the work of Yuan et al\cite{Yuan2019} by matching the entire concurrent RCT rather than only a subset of the active treatment arm for the following rationales. Randomly subsampling matches from active treatment group for concurrent control and then obtaining matches from external control for the rest of the treated\textit{ (unmatched treated)} is equivalent to randomly selecting a subset from treatment arm directly used for obtaining matches from external control. Such randomly selected subset may not be representative of the RCT population, especially when it is selected only for once and the sample size is not large. To alleviate this concern, Yuan at el.\cite{Yuan2019} proposed to repeat the procedure for multiple times and use the average of the means of the multiple matched subsets to derive the estimate of the mean outcome under the concurrent control arm. However, as a consequence, its variance becomes more complicated. Yuan et al.\cite{Yuan2019} proposed to use the average of the estimated variances over different repetitions as the variance estimate, but this approach provides overestimation in general, as the multiple matched sets are very unlikely to completely overlap. More details will be discussed in Section \ref{sec2} and numerical justification will be provided in Section \ref{sec4}. By matching the entire concurrent RCT, we can easily take the full advantages of matching \textit{without} replacement, such as the downstream diagnosis and analysis. Estimator can be formed by a weighted average of the concurrent and matched external controls, whose variance can be estimated by simply assuming the data after matching (\textit{without} replacement) are approximately independent \cite{Schafer2008}, or by bootstrapping the matched pairs or sets \cite{Austin2014}. Moreover, an RCT is often associated with multiple treatment arms, where a direct extension of the framework proposed by Yuan at el. \cite{Yuan2019} would construct different sets of matches from external control for each active treatment arm. By using our proposed approach, a common set of control subjects is used for comparison, which is more similar to what a typical RCT does. Additionally, our proposed method does not require unblinding the treatment assignment for matching during the trial execution, which helps to maintain the data integrity and credibility in the analysis of RCT. Therefore, matching can be properly diagnosed and fine-tuned and the matched set can be fixed before the trial unblinding.

The rest of this article proceeds as follows. We introduce notations based on potential outcome framework and review the NC matching proposed by Yuan at el. \cite{Yuan2019} in Section \ref{sec2}. In Section \ref{sec3}, we develop our matching design, introduce the estimator and a bootstrap method for variance estimation that was introduced in the paper written by Austin and Small \cite{Austin2014}. In section \ref{sec4}, real data based simulation studies are conducted to evaluate and illustrate our method. Section \ref{sec5} provides the summary.

\section{Preliminaries}\label{sec2}

\subsection{Framework}
Suppose we have an RCT with $k$ active treatment (AT) arms and one concurrent control (CC) arm (i.e., placebo), we assume the sample size of CC is less than that of the AT arms, and we use matched external control (EC) samples to augment the CC arm. 
Let $D$ denote the data source, i.e., $D=1$ if patient comes from RCT and $D=0$ from EC. Let $A$ be the treatment indicator that equals 0 if the subject is assigned with placebo (either in RCT or EC) and equals $a$ if the patient is assigned to the $a$-th AT arm. Let $Y^{(0)}$ and $Y^{(a)}$ be the potential outcomes under control and treatment $a$, respectively, where $a=1,...,k$. For patient $i$, let $D_i$, $A_i$ and $Y_i^{(a)}$ $(a=0,1,...,k)$, be realizations of $D$, $A$ and $Y^{(a)}$, respectively, where $i=1,...,n_r+n_e$, $n_r=n_0+n_1+\cdots+n_k$ with $n_a$ being the sample size of treatment $a$ in RCT $(a=0,1,...,k)$, and $n_e$ is the total number of matched samples from EC. We observe $Y_i=Y_i^{(a)}$ if and only if $A_i=a$.

Since EC is used only for augmenting the RCT, our primary estimand of interest stays unchanged from the estimand of RCT, namely, the average treatment effect (ATE) under the RCT population, denoted by
$$\theta_a = E(Y^{(a)}\mid D=1) - E(Y^{(0)}\mid D=1),$$
where $a=1,...,k$. Apparently, $\theta_a$ can be estimated by using RCT data only. However, if we assume that RCT and EC are unconfounded (Assumption 1) and that the support of the observed baseline covariates $X$ for the RCT is a subset of the support of $X$ for the EC (Assumption 2), the EC data can be used to identify $E(Y^{(0)}\mid D=1)$, hence, to augment the CC in the RCT.
\begin{itemize}
    \item \textbf{Assumption 1 (unconfoundedness).} $Y^{(0)} \perp D\mid X$;
    \item \textbf{Assumption 2 (overlap).} $\text{Pr}(D=1\mid X)<1-\eta$ for some $\eta>0$.
\end{itemize}

The above two assumptions are well known for identifying the "average treatment effect on the treated (ATT)" in observational studies. Assumption 1 is sometimes also termed as “ignorable" or “no hidden bias”. See the nice review paper written by Imbens \cite{Imbens2004} for more details. 
Therefore, many existing techniques for identifying ATT can be leveraged to identify $E(Y^{(0)}\mid D=1)$ using the EC data, among which we focus on matching \textit{without} replacement \cite{Rosenbaum1985, Gu1993, Rosenbaum2020} in this article, which was also the approach taken by Yuan at el. \cite{Yuan2019}.

\subsection{Review of NC matching}
Yuan et al.\cite{Yuan2019} considered an RCT with only two arms, one AT arm ($k=1$) with sample size $n_1$ and one placebo arm with sample size $n_0$, where $n_1>n_0$.
In an effort to mimic 1:1 randomization, they proposed to obtain $n_1-n_0$ matched control samples from the EC via matching \textit{without} replacement. In detail, first $n_0$ subjects are randomly selected from the AT arm, which are considered as matches for the $n_0$ subjects in the CC arm. This is because by randomization all the measured and unmeasured covariates between treatment and control should be balanced. Then, the rest $n_1-n_0$ treated subjects in the AT arm are matched from the EC data by matching \textit{without} replacement on the estimated propensity score \cite{Rosenbaum1985}.
The resulting point estimator for $\theta_1$ is
\begin{equation}\label{eq1}
\tilde\delta_1 = \bar Y_{1} - \frac{n_0\bar Y_0 + (n_1-n_0)\tilde Y_e}{n_1},
\end{equation}
where $\bar Y_{1}$ and $\bar Y_0$ are the sample means of the AT and CC arms of the RCT, respectively, and $\tilde Y_e$ is the sample mean of the outcome from the matched EC subjects. Although there may be weak dependence between matched samples from the EC and the RCT data caused by the matching step, in general, the matched samples are assumed to be approximately independent \cite{Schafer2008} with the RCT data for the matching \textit{without} replacement. Yuan et al.\cite{Yuan2019} also imposed this assumption. 

In fact, this matching procedure is equivalent to randomly selecting $n_t-n_0$ treated subjects to be directly matched with EC data. However, these randomly selected $n_t-n_0$ treated subjects may not be representative of the RCT population. To overcome this shortcoming, Yuan et al.\cite{Yuan2019} proposed to repeat the procedure $J$ (>1) times and use the average of the estimators from each repetition to serve as the estimator of $\theta_1$, which can be simplified as
\begin{equation}\label{eq2}
\hat\delta_1 = \bar Y_{1} - \frac{n_0\bar Y_0 + (n_1-n_0)\hat Y_e}{n_1}, 
\end{equation}
where $\hat Y_e = J^{-1}\sum_{j=1}^{J}\tilde Y_{ej}$ and $\tilde Y_{ej}$ is the sample mean of the outcome from the $n_1-n_0$ matched subjects in the EC data of the $j$-th repetition. To estimate the variance of $\hat\delta_1$, Yuan et al.\cite{Yuan2019} proposed to simply take the average of the variance estimates from each of the $J$ repetitions. It is not difficult to show that this average can overestimate the variance as the $J$ matched sets for constructing $\tilde Y_{ej}$ ( $j=1,...J$) are likely to have distinct samples. A simple analogous example could be that $\left[Var\left(\frac{Z_1+Z_2}{2}\right)+Var\left(\frac{Z_2+Z_3}{2}\right)\right]/2\geq Var\left(\frac{(Z_1+Z_2)/2+(Z_2+Z_3)/2}{2}\right)$ when $Z_1, Z_2$ and $Z_3$ are independently and identically distributed. The simulation study in Section \ref{sec4} provides numerical justification.

In addition, RCT is often associated with multiple treatment arms (>2), while generalizing the framework of Yuan et al.\cite{Yuan2019} to an RCT with such setting is not straightforward. A direct extension is to apply NC matching to each of the AT arms. However, in this way, each AT arm from the same RCT is compared with a different set of control subjects, which the trialists and regulatory bodies may find difficult to comprehend. Furthermore, another limitation of the NC matching is that it can only be performed after unblinding the trial when the information on the treatment assignment can be accessed. This may limit the flexibility to deal with potential issues that may arise during matching.

\section{Match the entire concurrent RCT}\label{sec3}

In light of the aforementioned concerns, we propose to match the entire concurrent RCT rather than only a randomly selected subset of the treated subjects. Since the matches from the EC data are used to augment the estimate of $E(Y^{(0)}\mid D=1)$, utilizing all subjects from RCT as reference for obtaining matches from EC data can make the matches as comparable as possible to the RCT population ($D=1$). 
Additionally, matching the entire RCT can be applied to an RCT with multiple arms, where every AT arm of the RCT is compared with the same set of control subjects. A weighted estimator similar to $\hat \delta_1$ can be constructed to resemble the balanced allocation design, which is 
\begin{equation}\label{eq3}
\bar\delta_a = \bar Y_{a} - \frac{n_0\bar Y_0 + (n_{a}-n_0)\bar Y_e}{n_{a}}, 
\end{equation}
where $\bar Y_e$ is the sample mean of the outcome from the matched EC subjects and $n_a$ is the sample size for AT arm $a$, $a=1,..,k$. In fact, $\bar\delta_a$ can be written as a general weighted estimator,
\begin{equation}\label{eq4}
\bar\delta_a = \bar Y_{a} - \left[w\bar Y_0 + (1-w)\bar Y_e\right],
\end{equation}
where $w\in(0,1)$. Basically, $w$ determines how much contribution the CC data makes to the estimation of $E(Y^{(0)}\mid D=1)$. For example, the three estimators given by \eqref{eq1}, \eqref{eq2} and \eqref{eq3} all have $w=n_0/n_1$ which is motivated by the balanced allocation design if the CC is augmented by the subjects matched from EC. The general form \eqref{eq4} of $\bar \delta_a$ is also useful in the situation where an RCT has multiple different sized AT arms. For example, in a dose-ranging Ph2b study, the higher doses may have larger sample size than the lower doses. Regardless of the number of the AT arms, the augmented control group remains the same, which is an important feature when we aim to mimic an RCT. We recommend to fix $w$ before matching is conducted, or at least $w$ should not be driven by the outcome of the matched subjects. Otherwise $w$ is no longer a constant but a variable, whose variability should be accounted for in the variance estimation of $\bar \delta_a$.
It is a common choice to determine $w$ by sample sizes $n_a$, $a=0,...,1$.

Another advantage of this proposed approach is the ability to fine-tune the matching while the trial is ongoing without
sacrificing the integrity of the data and the credibility of the analysis. Given that our proposed approach does not require the
information on treatment assignments, matching can be performed once the trial completes the enrollment. By conducting matching (\textit{without} replacement) for multiple times with varied algorithms and tools, the matching result can be carefully diagnosed such that the matched EC set is as comparable as possible to the RCT population. If the EC has sufficient large samples overlapping with the RCT, 1:N matching may also be considered. Please refer to Chapter 8, part II of the book written by Rosenbaum \cite{Rosenbaum2020} for detailed discussion about matching \textit{without} replacement.

Regarding the statistical inference, one simple approach is to follow Schafer and Kang \cite{Schafer2008} that assumes the matched EC samples are approximately independent, under which the variance estimation is straightforward and the corresponding standard error (SE) is given by $SE(\bar\delta_a) = \sqrt{n_a^{-1}S_a^2 + \left[n_0^{-1}w^2+n_e^{-1}(1-w)^2\right]S_0^2}$, where $S_a^2$ is the sample variance of $Y_i$'s from the subjects in the AT arm $a$, and $S_0^2$ is the sample variance of $Y_i$'s from the subjects in the control group combining the CC arm and the set matched from the EC data. 
Another approach is to follow Austin and Small \cite{Austin2014} that bootstraps the matched pairs formed by matching \textit{without} replacement, under which the standard deviation (SD) of the estimators across all bootstrap samples is used as the SE for $\bar\delta_a$. 
The 95\% confidence interval constructed by $\bar\delta_a \pm 1.96 SE(\bar\delta_a)$ can be applied to both approaches. Our simulation in Section \ref{sec4} shows that the simple approach controls type-I-error well under the null. However, it may be slightly conservative under the alternative { in that} the simple approach provides overestimation of the variance due to ignoring the weak dependence caused by matching. On the other hand, the bootstrap approach performs better in terms of both well-controlled type-I-error and adequate coverage probability as bootstrapping matched pairs accounts for the within pair dependence caused by matching.

\section{Real data based simulation}\label{sec4}

We perform the simulations based on a real clinical trial in diabetes therapeutic area to  evaluate the new matching design for augmenting an RCT with external control in terms of the bias and variance estimation of the associated estimator. Comparison with the NC matching is also studied. Since our goal is not to examine different algorithms or tools for matching \textit{without} replacement, the commonly used optimal matching on the estimated propensity score \cite{Gu1993, Rosenbaum1985, Rosenbaum2020} is applied and conducted by function \textit{"pairmatch"} from the R-package \textit{"optmatch"} throughout all simulations.

\subsection{Settings and data-generating process}

The real clinical trial consists of patients who completed the 52-week measurement of HbA1c, a metric used to evaluate the patients' level of blood glucose. The original change from baseline in HbA1c observed from the study serves as the potential outcome $Y^{(0)}$ under placebo, for which smaller value indicates more improvement. We use five baseline covariates, including baseline HbA1c, BMI, waist circumference, gender and age, denoted as $\bfX=(X_1,..., X_5)$. 
The empirical distribution of $(Y^{(0)}, \bfX)$ of all patients serves as super-population distribution in simulations. We considered three different settings for potential outcomes under the active treatment arms,
\begin{itemize}
    \item Setting 1: $Y^{(1)}=Y^{(0)}+\epsilon$ with $\epsilon ~\sim N(0,0.5)$ and $\epsilon \perp (Y^{(0)}, \bfX)$;
    \item Setting 2: $Y^{(1)}=Y^{(0)}-1$;
    \item Setting 3: $Y^{(1)}=Y^{(0)}-1$ and $Y^{(2)}=Y^{(0)}-0.2X_1$.
\end{itemize}
Note that the true relationship between $Y^{(0)}$ and $\bfX$ is unknown as it comes from a real clinical trial, neither is that between $Y^{(1)}, Y^{(2)}$ and $\bfX$. Those unknown relationships are retained in the real data based simulations.

For each subject drawn from the super-population, the probability of being selected to the RCT or EC is determined by the following logistic propensity score model: 
\begin{equation}\label{ps}
logit\{\text{Pr}(D=1 | X)\} = \alpha + X_1+0.05X_2+0.01X_3+0.4X_4-0.02X_5,
\end{equation}
where $\alpha$ is selected to control the ratio of the numbers of the subjects in the RCT versus EC to be approximately 1:10, 
and the coefficients for $\bfX$ are determined to differentiate the baseline covariates distribution between the RCT and EC.

For the RCT in setting 1 and 2, we consider 2:1 allocation ratio for the active treatment arm and the placebo arm, and two different total sample sizes ($n_{r}$) of 90 and 180. In setting 3 where the RCT has more than two arms, we consider a 2:2:1 allocation ratio among AT arm 1, AT arm 2 and the placebo arm with total sample sizes being 150 or 300. In each of the settings, there are 30 or 60 subjects assigned to the placebo arm (CC) of the RCT. The data-generating proceeds as follows for each simulation:
\begin{itemize}
    \item Randomly draw a sample $i$ from the super-population distribution of $(Y^{(0)},Y^{(1)},Y^{(2)},\bfX)$, which has covariates $\bfX_i$ and potential outcomes $(Y^{(0)}_i,Y^{(1)}_i,Y^{(2)}_i)$. 
    \item Randomly draw a data source indicator $D_i$ for sample $i$ based on a Bernoulli distribution with probability $\text{Pr}(D_i=1|\bfX_i)$ computed according to the propensity score model \eqref{ps}. If $D_i=1$, sample $i$ belongs to the RCT, otherwise it belongs to the EC.
    \item Repeat step 1 and 2 until we have $n_r$ total subjects in the RCT. With carefully selected $\alpha$ in the propensity score model \eqref{ps}, we expect to have approximately $10n_r$ subjects in the EC.
    \item Randomly assign treatment $A \in \{0,1\}$ (or $A \in \{0,1,2\}$)  to the $n_r$ subjects in the RCT with a fixed treatment allocation ratio 2:1 (or 2:2:1). 
    \item The observed outcome for each subject $i$ is $Y_i=D_i\left(Y_i^{(0)}I_{\{A_i=0\}}+Y_i^{(1)}I_{\{A_i=1\}}+Y_i^{(2)}I_{\{A_i=2\}}\right)+(1-D_i)Y_i^{(0)}$.
\end{itemize}

\subsection{Analyses and results}
For each simulated data set, we first estimate the propensity score for being selected to the RCT or EC using a logistic regression which regresses data source indicator $D$ on the observed baseline covariates $\bfX$. We then apply the optimal matching on the estimated propensity score \cite{Gu1993, Rosenbaum1985, Rosenbaum2020} to obtain a matched set from the EC data for the entire RCT without replacement. The equation \eqref{eq4} with $w=0.5$ is used to estimate $\theta_a$, which is equivalent to use equation \eqref{eq3} as in all settings $n_0/n_a = 0.5$. The standard error is evaluated with both simple and bootstrap approaches discussed in Section \eqref{sec3}. For the simple approach, all subjects in the RCT and matched EC are assume to be approximately independent, while for the bootstrap approach, matched pairs formed by matching are drawn 500 times with replacement to account for the within-pair dependence.

NC matching is studied in setting 1 and 2 where RCT only has two arms. We repeat NC matching for different times, where $J=1, 3$ and $5$. When $J>1$, we also report the average count of distinct EC subjects used to construct $\hat Y_e$ in formula \eqref{eq2}. This number is usually between $n_1-n_0$ and $J\times(n_1-n_0)$ as multiple matched EC sets are very likely to { share common} subjects. The analysis based only on the RCT data is provided as a reference for all settings. The first setting is intended to examine the control of Type-I-error as $\theta_1 =  E(Y^{(1)}\mid D=1) - E(Y^{(0)}\mid D=1)=0$. The second setting is to assess the performance of the varied methods under a case where there is beneficial treatment effect, i.e., a 1\% improvement in mean HbA1c change at week 52 comparing the treatment arm to the placebo arm. The last setting considers an RCT with more than two arms aiming to show the flexibility of the proposed matching design.

Based on 5000 simulations, we compute the average of the bias, the standard deviation (SD), and the average of the standard error (SE), of the estimators under different settings and methods. For Setting 1, Type-I-Error is provided, and for Setting 2 and 3, coverage probability (CP) of 95\% confidence interval is shown. The simulation results for Setting 1 and 2 and Setting 3 are summarized in Table 1 and Table 2, respectively. 

All methods have negligible bias for all settings and both NC matching and our proposed method have smaller SD than the raw method that only uses the RCT data, where the benefit of leveraging the external information is more than 20\% in terms of efficiency gain. According to the results of Setting 1, all methods control the type-I-error well, but it appears that NC matching has type-I-error much smaller than 0.05 and correspondingly its SE overestimates the SD, which can also be seen from the results of Setting 2. In addition, as the number of repetitions of NC matching increases, the SD is decreasing due to the more distinct matches used in $\hat Y_e$, however, the SE remains similar. Hence, the overestimation becomes more apparent as $J$ increases. For our proposed method, SEs are in overall close to SDs. Although the simple SE slightly overestimates SD in Setting 2, the bootstrap SE approximates SD well in all settings.

\section{Discussion}\label{sec5}
The use of information from real world to assess the effectiveness of medical products is becoming increasingly popular and more acceptable by regulatory agencies. In this paper, we have reviewed the NC matching proposed by Yuan et al.\cite{Yuan2019} for augmenting a randomized clinical trial with external control data. We further propose an improved matching design that matches the entire RCT  rather than a randomly selected subset. As such, the matched external control subjects are more comparable to the RCT population. Using a weighted estimator with a fixed weight, our method can be easily applied to an RCT with multiple treatment arms and every active treatment arm in the RCT is compared with a common control group. Moreover, we allow matching to be conducted before trial unblinding, i.e., without accessing the treatment assignment. This means, matching can be finely tuned using varied algorithms and tools, and carefully diagnosed while still maintaining the data integrity and { analysis credibility}. For the variance estimation of the weighted estimator, we suggest to bootstrap the matched pairs, inspired by Austin and Small \cite{Austin2014}, which shown in simulation controls type-I-error well and provides adequate coverage probability.

Despite the advantages of our proposed method over NC matching, both methods rely on the strong assumption of no unmeasured confounding bias, which may limit the usage and the reliability of the result. Under casual inference framework, one standard approach is to conduct sensitivity analysis to quantify the magnitude of the unmeasured confounding bias required to overturn the conclusion, which is also suggested by Rosenbaum \cite{Rosenbaum2020} who introduced the propensity score \cite{Rosenbaum1983} and developed throughout his career a series of matching methods and a comprehensive framework of sensitivity analysis for matched observational studies.

\section*{CONFLICT OF INTEREST}
The authors declare that there is no conflict of interest for this article.
\section*{DATA AVAILABILITY STATEMENT}
Data sharing is not applicable to this article as no new data were created or analyzed in this study.

\bibliography{wileyNJD-AMA}

\clearpage
\begin{table}[ht]
	\begin{center}
		\caption{Bias, standard deviation (SD), estimated SD (SE), and type-I-error or coverage probability (CP) for Setting 1 and 2} \vspace{-2mm}
		\resizebox{0.9\textwidth}{!}{
			\begin{tabular}{cccccccccccccc}
			\hline\\[-2.5ex]\hline\\[-2ex]
			&       &     &              &  & \multicolumn{4}{c}{Setting 1 ($\theta_1=0$)} &  & \multicolumn{4}{c}{Setting 2 ($\theta_1=-1$)}\\[0.5ex]
			\cline{6-9} \cline{11-14} \\[-2ex]
			$n_r$ & \multicolumn{3}{c}{Method} &  & Bias   & SD     & SE    & type-I-error &  & Bias    & SD   & SE     & CP    \\[0.5ex]
			\hline\\[-2ex]
			90    & \multicolumn{3}{c}{Raw}    &  & -0.003 & 0.321  & 0.319 & 0.058        &  & 0.000   & 0.303 & 0.305 & 0.950 \\[0.5ex]
			\cline{2-4}\\[-2ex]
			& \multicolumn{3}{c}{NC matching}  &  &        &        &       &              &  &          &      &       &       \\
			& J     &     & Distinct Matches   &  &        &        &       &              &  &          &      &       &       \\
			\cline{2-2} \cline{4-4} \\[-2ex]
			& 1     &     & 30.0               &  & -0.001 & 0.250  & 0.265 & 0.039        &  & 0.001    & 0.226 & 0.248 & 0.970 \\
			& 2     &     & 54.4               &  & 0.002  & 0.235  & 0.265 & 0.030        &  & 0.000    & 0.216 & 0.248 & 0.973 \\
			& 3     &     & 75.7               &  & 0.000  & 0.233  & 0.265 & 0.027        &  & 0.000    & 0.211 & 0.248 & 0.979 \\[0.5ex]
			\cline{2-4}\\[-2ex]
			& \multicolumn{3}{c}{New matching design} &  & &        &       &              &  &          &       &       &       \\
      & \multicolumn{3}{c}{simple SE}   &  & \multirow{2}{*}{0.001} & \multirow{2}{*}{0.241} & 0.245 & 0.047        &  & \multirow{2}{*}{0.003} & \multirow{2}{*}{0.218} & 0.227 & 0.959 \\
      & \multicolumn{3}{c}{bootstrap SE} &  &                       &                        & 0.238 & 0.058        &  &                        &                        & 0.220 & 0.952 \\[0.5ex]
      \hline \\[-2ex]
      180   & \multicolumn{3}{c}{Raw}          &  & 0.002  & 0.230  & 0.225  & 0.054        &  & -0.005   & 0.218 & 0.216 & 0.948 \\[0.5ex]
      \cline{2-4} \\[-2ex]
      & \multicolumn{3}{c}{NC matching}  &  &  &                    &       &              &  &     &  &       &       \\
      & J     &     & Distinct Matches   &  &  &                    &       &              &  &     &  &       &       \\
      \cline{2-2} \cline{4-4} \\[-2ex]
      & 1     &     & 60.0               &  & 0.004                  & 0.176                  & 0.187 & 0.035        &  & -0.001                 & 0.160                  & 0.176 & 0.971 \\
      & 2     &     & 108.1              &  & 0.003                  & 0.167                  & 0.187 & 0.028        &  & -0.002                 & 0.153                  & 0.176 & 0.974 \\
      & 3     &     & 149.1              &  & 0.004                  & 0.165                  & 0.187 & 0.025        &  & 0.000                  & 0.149                  & 0.176 & 0.980 \\[0.5ex]
      \cline{2-4} \\[-2ex]
      & \multicolumn{3}{c}{New matching design} &  &                        &                        &       &              &  &                        &                        &       &       \\
      & \multicolumn{3}{c}{simple SE}    &  & \multirow{2}{*}{0.005} & \multirow{2}{*}{0.172} & 0.173 & 0.049        &  & \multirow{2}{*}{0.001} & \multirow{2}{*}{0.156} & 0.161 & 0.957 \\
      & \multicolumn{3}{c}{bootstrap SE} &  &                        &                        & 0.168 & 0.057        &  &                        &                        & 0.155 & 0.949 \\[0.5ex]
      \hline
\end{tabular}}
	\end{center}
\end{table}

\begin{table}[ht!]
	\begin{center}
		\caption{Bias, standard deviation (SD), average estimated SD (SE), and coverage probability (CP) for Setting 3} \vspace{-2mm}
		\resizebox{0.8\textwidth}{!}{
			\begin{tabular}{cccccccccccc}
	  \hline\\[-2.5ex]\hline\\[-2ex]
	  &              &  & \multicolumn{4}{c}{$\theta_1=-1$}                               &  & \multicolumn{4}{c}{$\theta_2=-1.651$}                          \\[0.5ex] \cline{4-7} \cline{9-12} \\[-2ex]
$n_r$ & Method       &  & Bias                   & SD                     & SE    & CP    &  & Bias                   & SD                     & SE    & CP    \\[0.5ex]
      \hline\\[-2ex]
150   & Raw          &  & -0.006                 & 0.309                  & 0.306 & 0.941 &  & -0.005                 & 0.305                  & 0.300 & 0.943 \\[0.5ex]
      \cline{2-2} \\[-2ex]
      & New matching &  &                        &                        &       &       &  &                        &                        &       &       \\
      & simple SE    &  & \multirow{2}{*}{0.000} & \multirow{2}{*}{0.220} & 0.223 & 0.952 &  & \multirow{2}{*}{0.001} & \multirow{2}{*}{0.214} & 0.215 & 0.954 \\
      & bootstrap SE &  &                        &                        & 0.218 & 0.946 &  &                        &                        & 0.212 & 0.946 \\[0.5ex]
      \hline\\[-2ex]
300   & Raw          &  & -0.002                 & 0.218                  & 0.215 & 0.944 &  & -0.002                 & 0.212                  & 0.211 & 0.947 \\[0.5ex]
      \cline{2-2}\\[-2ex]
      & New matching &  &                        &                        &       &       &  &                        &                        &       &       \\
      & simple SE    &  & \multirow{2}{*}{0.002} & \multirow{2}{*}{0.155} & 0.157 & 0.950 &  & \multirow{2}{*}{0.002} & \multirow{2}{*}{0.150} & 0.152 & 0.952 \\
      & bootstrap SE &  &                        &                        & 0.153 & 0.943 &  &                        &                        & 0.149 & 0.945 \\[0.5ex]
      \hline
\end{tabular}}
	\end{center}
\end{table}


\end{document}